\documentclass[pdflatex,sn-mathphys]{sn-jnl}% Math and Physical Sciences Reference Style
\jyear{2023}%

\theoremstyle{thmstyleone}%
\theoremstyle{thmstyletwo}%

\theoremstyle{thmstylethree}%

\begin{document}

\title[Superconducting Arcs]{Superconducting Arcs}

\author[1]{\fnm{Andrii} \sur{Kuibarov}}
\author[1,2]{\fnm{Oleksandr} \sur{Suvorov}}
\author[1]{\fnm{Riccardo} \sur{Vocaturo}}
\author[1,3]{\fnm{Alexander} \sur{Fedorov}}
\author[1,3]{\fnm{Rui} \sur{Lou}}
\author[1]{\fnm{Luise} \sur{Merkwitz}}
\author[3]{\fnm{Vladimir} \sur{Voroshnin}}
\author[4]{\fnm{Jorge I.} \sur{Facio}}
\author[1]{\fnm{Klaus} \sur{Koepernik}}
\author[5]{\fnm{Alexander} \sur{Yaresko}}
\author[1]{\fnm{Grigoriy} \sur{Shipunov}}
\author[1]{\fnm{Saicharan} \sur{Aswartham}}
\author[1]{\fnm{Jeroen} \sur{van den Brink}}
\author[1]{\fnm{Bernd} \sur{B\"uchner}}

\author*[1]{\fnm{Sergey} \sur{Borisenko}}\email{S.Borisenko@ifw-dresden.de}

\affil[1]{\orgname{Leibniz Institute for Solid State and Materials Research}, \orgaddress{\street{Helmholzstraße 20}, \city{Dresden}, \postcode{01069}, \country{Germany}}}

\affil[2]{\orgname{Kyiv Academic University}, \orgaddress{\street{36 Vernadsky blvd}, \city{Kyiv}, \postcode{03142}, \country{Ukraine}}}

\affil[3]{\orgname{Helmholtz-Zentrum Berlin f\"ur Materialien und Energie}, \orgaddress{\street{BESSY II}, \city{Berlin}, \postcode{12489}, \country{Germany}}}

\affil[4]{\orgname{Centro Atómico Bariloche, Instituto de Nanociencia y Nanotecnología
(CNEA-CONICET) and Instituto Balseiro}, \orgaddress{\street{Av. Bustillo},  \postcode{9500},  \country{Argentina}}}

\affil[5]{\orgname{Max Planck Institute for Solid State Research}, \orgaddress{\street{Heisenbergstraße 1}, \city{Stuttgart}, \postcode{70569},  \country{Germany}}}

%\abstract{Topological superconductivity \cite{Sato_2017,Sharma_2022} is essential to generate Majorana fermions needed for quantum computing. Bulk topological superconductors remain elusive and two most promising approaches exploiting the proximity-induced superconductivity \cite{Proximity_Fu} seem difficult to realize  \cite{NanoWires_Mourik,Frolov,YSRIBS_Wang, BorisenkoNPJQM}.  Weyl semimetals due to their intrinsic topology \cite{Vishwanath} belong to potential candidates too \cite{Sato_2017,Sharma_2022}, but search for Majorana fermions has always been connected with the superconductivity in the bulk, leaving the possibility of intrinsic superconductivity of the Fermi surface arcs themselves practically without attention, even from the theory side.
%
%Here, by means of angle-resolved photoemission spectroscopy and ab-initio calculations, we unambiguously identify topological Fermi arcs on two opposing surfaces of the non-centrosymmetric Weyl material PtBi$_2$ \cite{Veyrat_Nanoletters}. We show that these states become superconducting at different temperatures around 10K. Remarkably, the corresponding coherence peaks appear as the strongest and sharpest excitations ever detected by photoemission from solids, suggesting significant technological relevance. Our findings indicate that topological superconductivity in PtBi$_2$ occurs exclusively at the surface, which not only makes it an ideal platform to host Majorana fermions, but may  also lead to a unique quantum phase - an intrinsic topological SNS Josephson junction. }

\abstract{An essential ingredient for the production of Majorana fermions that can be used for quantum computing is the presence of topological superconductivity \cite{Sato_2017,Sharma_2022}. As bulk topological superconductors remain elusive, the most promising approaches exploit proximity-induced superconductivity \cite{Proximity_Fu} making systems fragile and difficult to realize  \cite{NanoWires_Mourik,Frolov,YSRIBS_Wang, BorisenkoNPJQM}. 
Weyl semimetals due to their intrinsic topology \cite{Vishwanath} belong to potential candidates too \cite{Sato_2017,Sharma_2022}, but search for Majorana fermions has always been connected with the superconductivity in the bulk, leaving the possibility of intrinsic superconductivity of the Fermi surface arcs themselves practically without attention, even from the theory side.

Here, by means of angle-resolved photoemission spectroscopy and ab-initio calculations, we unambiguously identify topological Fermi arcs on two opposing surfaces of the non-centrosymmetric Weyl material PtBi$_2$ \cite{Veyrat_Nanoletters}. We show that these states become superconducting at different temperatures around 10K. Remarkably, the corresponding coherence peaks appear as the strongest and sharpest excitations ever detected by photoemission from solids, suggesting significant technological relevance. Our findings indicate that topological superconductivity in PtBi$_2$ occurs exclusively at the surface, which not only makes it an ideal platform to host Majorana fermions, but may  also lead to a unique quantum phase - an intrinsic topological SNS Josephson junction. }


\maketitle

\section{Introduction}\label{sec1}
\begin{figure}[h]%
\centering
\includegraphics[width=0.9\textwidth]{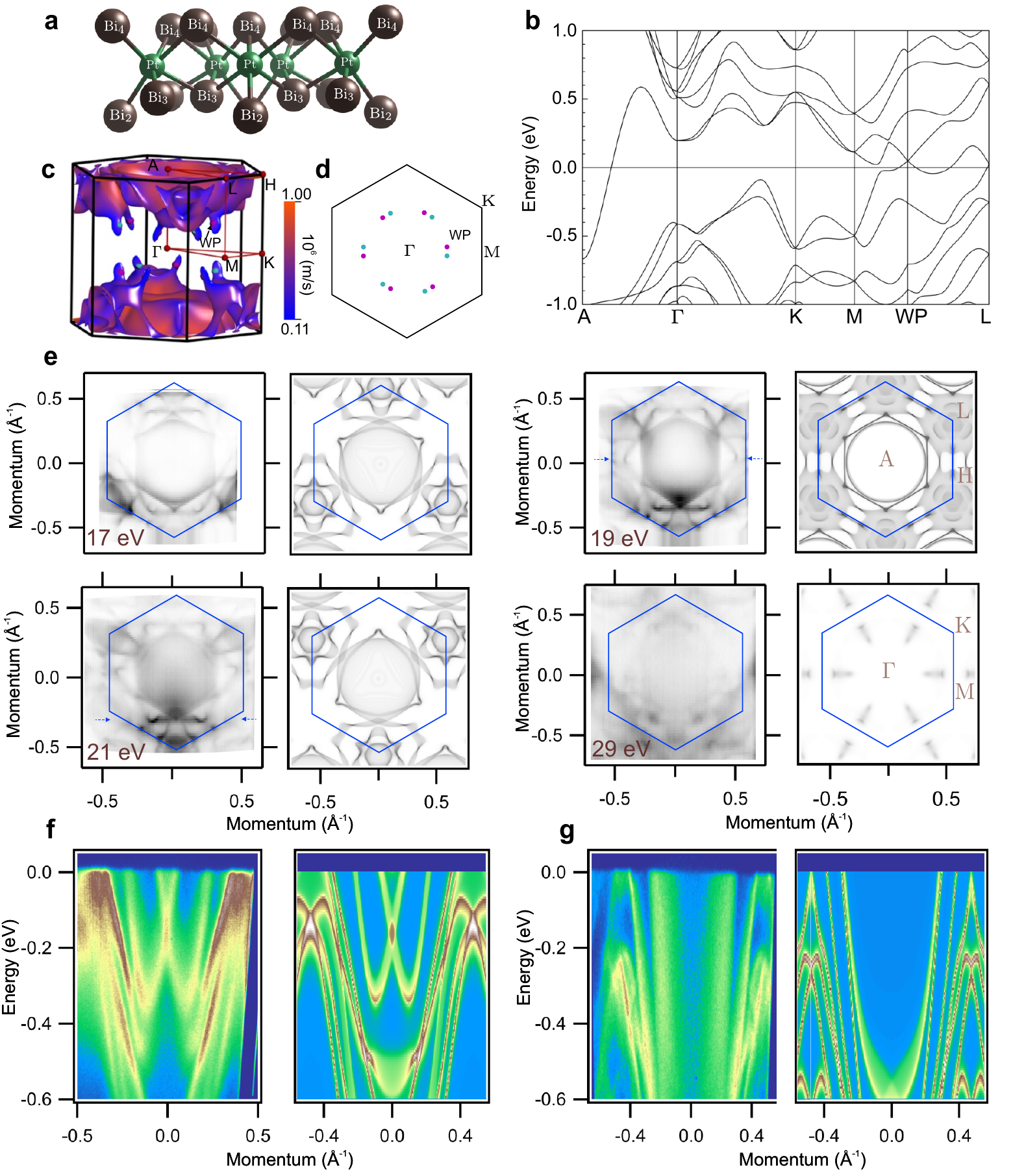}
\caption{3D Band Structure of PtBi$_2$. \textbf{a} Crystal structure of PtBi$_2$. \textbf{b} Fragment of the band structure. One Weyl point (WP) is included. \textbf{c} Fermi surface, Weyl and high symmetry points. Color scale corresponds to Fermi velocity. \textbf{d} $\Gamma$MK plane of the BZ zone with projections of the Weyl points. Magenta (blue) colors stand for positive (negative) chirality. \textbf{e} Fermi surface maps taken using different photon energies and corresponding results of the band structure calculations. We note, that fixed photon energy probes a sphere of the large radius in the k-space, matching theoretical data formally only at one point in the center. Theoretical Fermi maps were averaged of over a range of 1/10 of the BZ size in the $k_z$ direction to account for experimental uncertainties. The intensities of the theoretical Fermi maps were normalized to the DOS for different $k_z$ points. \textbf{f}, \textbf{g} left: Energy-momentum intensity distributions at 21 eV and 19 eV respectively along the cuts indicated by blue dashed arrows in panel \textbf{e}, right: corresponding energy-momentum spectra taken from the band structure calculation.}
\label{fig1}
\end{figure}
The realisation of topological superconductivity, which leads to robust Majorana fermions in new materials has so far been hindered by numerous experimental challenges. Among them are the sophisticated growth of nanowire single crystals and heterostructures as well as fine-tuning of the composition of non-stoichiometric compounds. Additionally, the rarity of spin-triplet  superconductors and extremely small inverted gaps in iron-based superconductors \cite{BorisenkoNPJQM}, proposed as intrinsic heterostructures \cite{Zhang2018}, are responsibe for the lack of success in already existing materials.

Weyl semimetals bear non-degenerate spin states both in the bulk and at the surface and the doped version of either time-reversal-breaking or noncentrosymmtric ones can become superconducting \cite{Meng_WeylSupc}. The search for topological superconductivity in such systems has been focused on finding bulk superconductivity which would lead to Majorana fermion surface states. The possibility of intrinsic superconductivity of the arcs themselves, related to topology of the bandstructure with Weyl nodes, has virtually not been considered. Although the arcs cannot support superconductivity in time-reversal-breaking Weyl semimetals \cite{Sato_2017}, the noncentrosymmetric ones remain an option.

Trigonal PtBi$_2$ has emerged recently as a type-I Weyl semimetal, which together with reported superconductivity \cite{Shipunov_PRM, Veyrat_Nanoletters} makes it an attractive candidate for topological superconductivity. Scanning tunneling spectroscopy (STM) experiments confirmed the presence of surface superconductivity by observing typical spectra of superconducting gaps \cite{Schimmel_gap, Shipunov_PRM}. Here we found that the topological Fermi surface arcs bear the superconductivity in PtBi$_2$. This occurs on both of the two nonequivalent surfaces of PtBi$_2$.

\section{Three-dimensional band structure}\label{sec2}

The electronic structure of trigonal PtBi$_2$ has been studied both experimentally and theoretically earlier \cite{YaoPRB2016,Setti,Jiang_JAP2020,Feng2019,Veyrat_Nanoletters,Gao2018}. The material crystallizes in $P31m$ space group \cite{Shipunov_PRM} and exposes two different surfaces upon cleaving, which we refer to as A and B below  (Fig. \ref{fig1}a). The band structure (Fig. \ref{fig1}b) arises mostly due to hybridization of Bi 6p, Pt 5d and Pt 6s states. Two sets of Weyl points are located in momentum space as shown in Fig. \ref{fig1}c, d having the energy of 47 meV above the Fermi level. In order to set a baseline from which the 3D band structure can be resolved we have recorded 16 ARPES datasets covering at least the first 3D Brillouin zone and approximately 1 eV in energy using the photon energies from 15 to 43 eV (see also Fig. S1 in SI). This allowed us to identify high-symmetry points along the $k_z$ direction and find the value of the inner potential ( $V_0$ = 10.5 eV ). In Fig. \ref{fig1}e we show the Fermi surface maps taken using the photon energies corresponding to high symmetry points along the $k_z$ direction and approximately half-way in between them. The latter two (left column) are easy to recognize due to pronounced $C_3$ symmetry of the pattern, rotated by 60$^{\circ}$ with respect to each other. The map taken with 19 eV photons has a higher degree of hexagonal shape, which corresponds to the A-point of the BZ. The map taken with 29 eV photons at the level of the $\Gamma$-point  bears a certain degree of resemblance with the almost featureless calculated intensity as suggested by Fig. \ref{fig1}c. The experimental pattern in this case is connected with the finite k$_z$ resolution of ARPES. In Fig. \ref{fig1}f, g we also show the comparison of the dispersions along the lines indicated in Fig. \ref{fig1}e. Although there is no one-to-one correspondence, the features look very similar being shifted in energy or momentum without any signatures of strong renormalization etc (see also Fig. S2 in SI). There are several reasons why the intensity distributions do not match exactly. First is that DFT calculations never perfectly reproduce the electronic structure, second is the influence of matrix elements on the photoemission intensity, third is the moderate $k_z$-resolution of ARPES and, finally, the band structure calculations in Fig. \ref{fig1} do not consider the presence of the surface. Nevertheless,the data presented in Fig. \ref{fig1}e-g suggest a reasonable general agreement between experiment and theory, which is in accord with all previous ARPES studies \cite{YaoPRB2016,Feng2019,Setti,Jiang_JAP2020}. The experimental confirmation of all the main features of the band structure near the Fermi level thus implies that PtBi$_2$ is indeed a Weyl semimetal, which we will fully confirm below.

\section{Surface states on two terminations}\label{sec3}
\begin{figure}[ht]%
\centering
\includegraphics[width=0.9\textwidth]{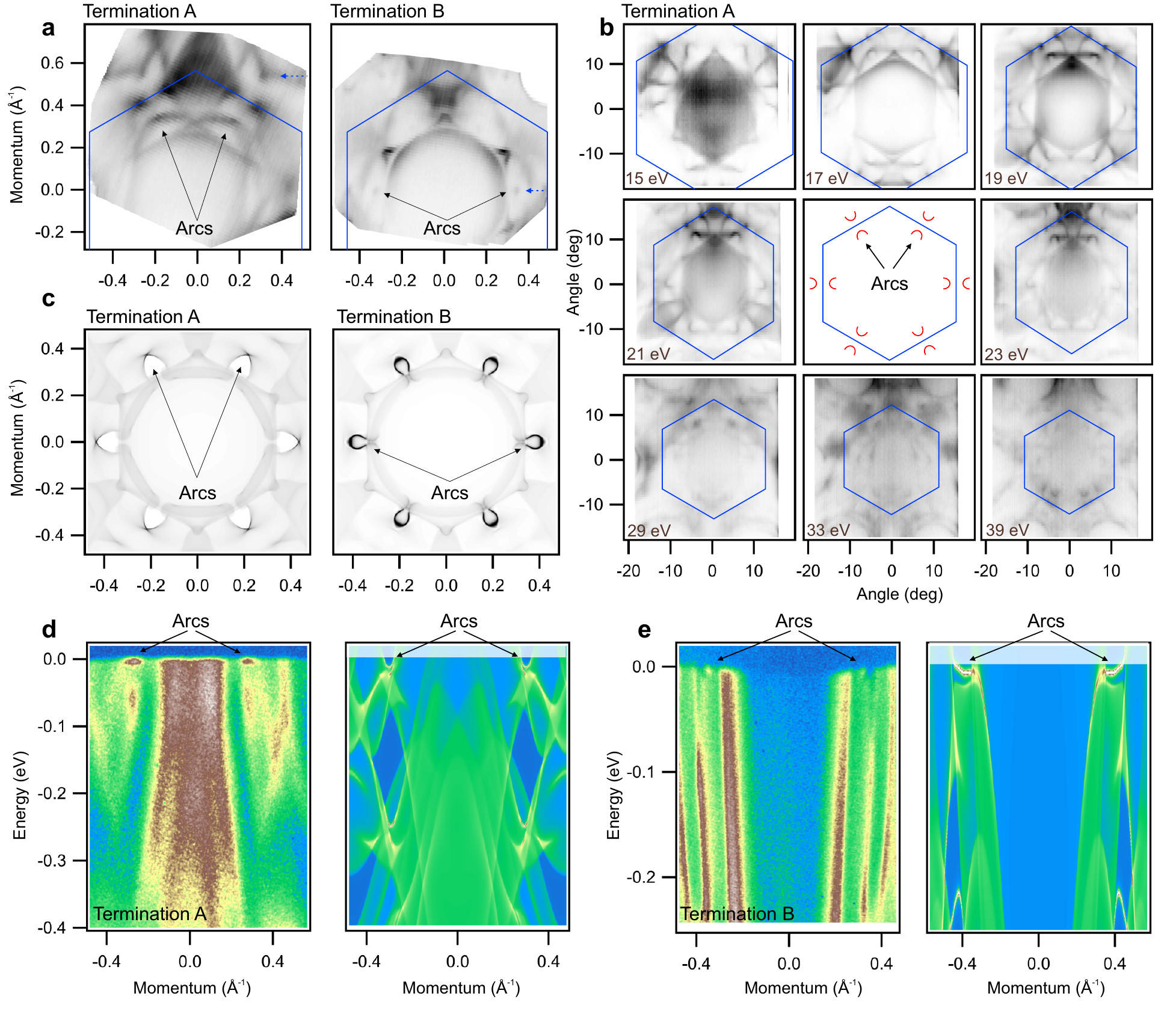}
\caption{Fermi surface arcs. \textbf{a} High-resolution Fermi surface maps ($h\nu$ = 17 eV, T = 1.5 K) from both terminations. Arcs in the first BZ are indicated by the arrows. Note their presence in the equivalent positions in the first and repeated BZ. \textbf{b} Fermi surface maps at different photon energies, all showing the presence of the arcs measured at 15K. The sketch in the middle is the guide to the eye. \textbf{c} Arcs as seen in the calculations. \textbf{d},\textbf{e} Energy-momentum intensity plots for terminations A and B along the cuts through the arcs highlighted by blue dashed arrows in panels \textbf{a}.   }
\label{fig2}
\end{figure}
In Fig. \ref{fig2}a we present high-resolution Fermi surface maps from both terminations. Although seemingly different, closer inspection suggests that they share mostly the same pattern provided intensity variations are taken into account. The number of localized features can be clearly distinguished in the map from termination A, at approximately 3/4 of $\Gamma$M distance and equivalent locations. These features have been overlooked in earlier ARPES studies \cite{YaoPRB2016,Feng2019,Setti,Jiang_JAP2020}. Since the calculated bulk continuum displayed in Fig. \ref{fig1}b does not contain any similar electronic states in this region, we consider those as originating from the surface. The termination B map also shows similarly located features, but they are more clearly seen in the second BZ. The underlying bulk-related intensity is higher, masking the presence of the surface states. In order to establish their presence unambiguously, we show eight Fermi surface maps taken using different photon energies in Fig. \ref{fig2}b. All maps exhibit all of the above-mentioned features at the same location - approximately 3/4 of $\Gamma$M distance, as in the case with termination A. Since it is unlikely that a particular bulk feature would be present in all of the recorded FS maps for a material with a highly three-dimensional electronic structure, we conclude that also these features represent the surface. The schematic plot (in the middle of Fig. \ref{fig2}b) summarizes our observations regarding the locations of the arcs made from considering the Fermi surface maps.

Detected spots of intensity, which we identified above as surface states, remarkably coincide with the results of calculations which take into account the presence of the surface (Fig. \ref{fig2}c). Since PtBi$_2$ is a Weyl semimetal, one does expect the presence of the topological Fermi arcs, different for terminations A and B. At the same time, in an idea type-I Weyl semimetal, the location of the starting and end points of the arcs should be identical as those are the projections of the Weyl points (see Fig. \ref{fig1}d). The Weyl points, non-degenerate crossings of the bands in three-dimensional k-space, are almost impossible to detect by ARPES directly because of the finite resolution, but the corresponding Fermi arcs have been repeatedly seen experimentally in various materials\cite{arcs1_Yang2015,arcs2_Lv,arcs3_Deng,arcs4_Eric,arcs5_Borisenko2019}. 

Fig. \ref{fig2}d,e demonstrates a comparison of the intensity distribution along the paths marked in Fig. \ref{fig2}a, which run through the arcs. The arcs are situated very close to the Fermi level and are well-distinguishable from the smeared-out by $k_z$-resolution bulk dispersions.

Taking into account the discrepancies in the experimental and theoretical 3D band structure (Fig. \ref{fig1}), we do not expect exact correspondence between the calculated Fermi arcs and the ARPES data, but the observed agreement proves not only that the experimental features are indeed the topological Fermi arcs, but also that PtBi$_2$ is a Weyl semimetal.

\section{Robust Fermi arcs from laser-ARPES }\label{sec4}
\begin{figure}[h]%
\centering
\includegraphics[width=0.9\textwidth]{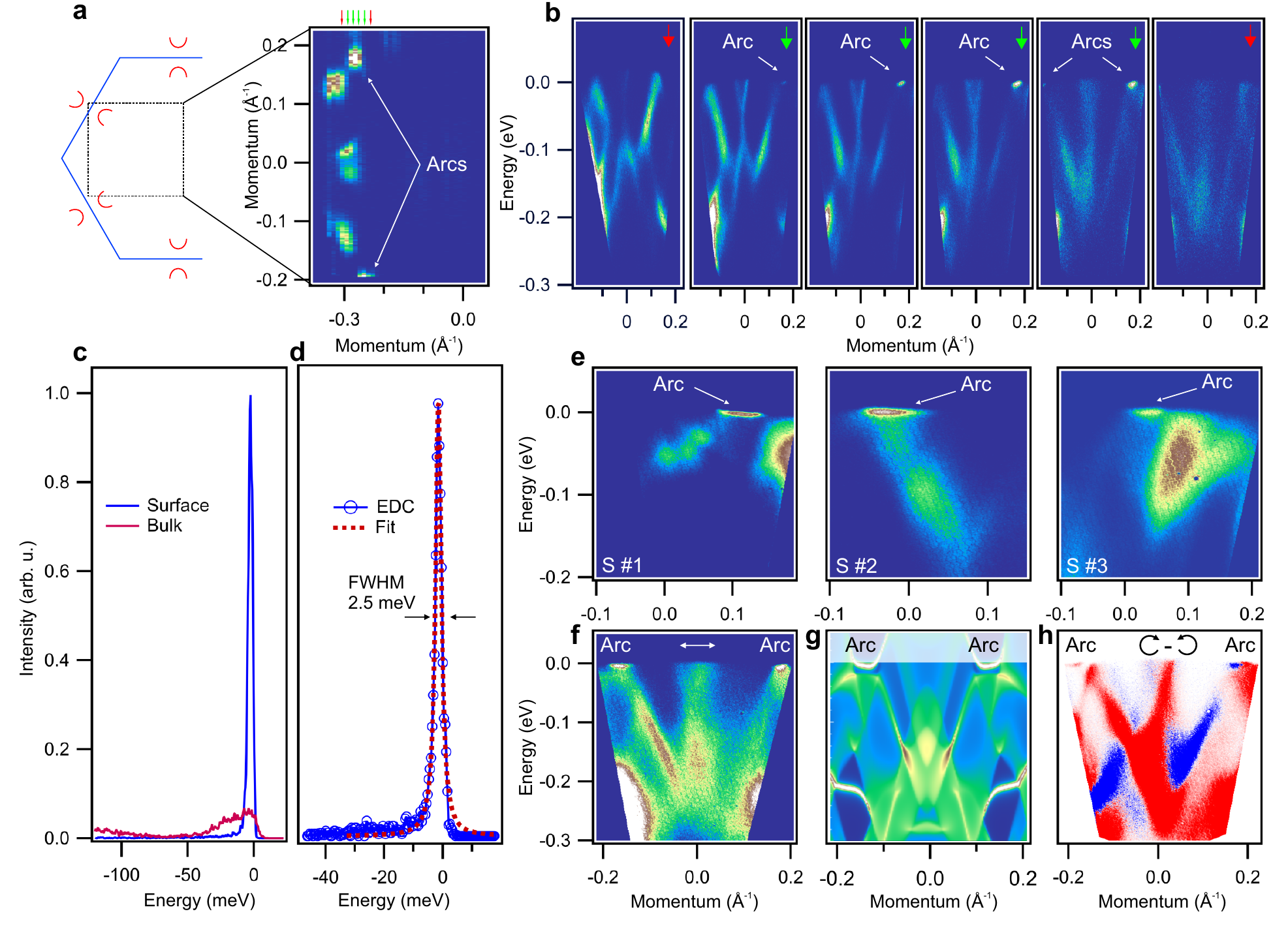}
\caption{Laser ARPES. \textbf{a} Fermi surface map taken using $h\nu$ = 5.9 eV at 3K. Arcs are seen together with other bulk-originated features. \textbf{b} Underlying dispersion along the momentum cuts indicated by arrows in \textbf{a}. \textbf{c} Typical energy distribution curves (EDC) from \textbf{b}. Bulk EDC is taken close to zeroth momentum, while surface EDC corresponds to the arc. \textbf{d} One of the narrowest and strongest EDCs detected in the present study. \textbf{e} Arcs seen along the different cuts through the BZ in different experimental geometries. \textbf{f} Intensity distribution taken using horizontally polarised light along the path crossing two arcs. \textbf{g} The same momentum and energy range as in f from the calculations. Note, that also the surface states at around 200 meV binding energies are reproduced. \textbf{h} Circular dichroism from the same region of the $k$-space.}
\label{fig3}
\end{figure}
In order to study the detected at the Fermi surface maps spots of intensity in more details, we carried out ARPES experiments using a laser setup. Because of the low kinetic energy of photoelectrons ( $\sim$ 1.7 eV), the part of the BZ accessible during these experiments is very limited. We have concentrated on detecting at least one of the arcs in the portion of the $k$-space marked in the sketch of Fig. \ref{fig3}a. Several representative cuts through the key features seen in the map are shown in Fig. \ref{fig3}b. Now the arc is better resolved, but still very localized in terms of both momentum and energy. We estimate the momentum extension to be of the order of 0.04 \AA$^{-1}$, which is in excellent agreement with theory (Fig. \ref{fig2}c).

The most striking characteristic of the arc states is their energy distribution. In Fig. \ref{fig3}c we compare the EDCs corresponding to the bulk and surface states. Since the data are taken with very high resolution and at extremely low temperature, the $k_F$-EDC representing bulk states has a well defined maximum (FWHM ~30 meV) near the Fermi level and leading edge width of approximately 5 meV. However, the sharpness and peak-to-background ratio of the EDC representing the surface arc is unprecedented. We have routinely observed the peaks having FWHM below 3 meV and $\sim$50 peak-to-background ratio in numerous cleaves of many samples (see Fig. S3 in SI). One of such curves is shown in Fig. \ref{fig3}d. To our best knowledge, such a sharp lineshape has never been observed in any photoemission experiment from solids earlier. 

In Fig. \ref{fig3}e we show further appearances of the arcs in the momentum-energy  plots from different cleaves and different terminations. The sharpness and flatness remain the robust characteristics of the feature in all our experiments at the lowest temperatures. We noticed that for A and B surfaces the arc states are supported by the strongly and weakly dispersing bulk states respectively, exactly as expected from theory.

Direct comparison with the calculations taking into account the presence of the topological surface states is presented in Fig. \ref{fig3}f,g. The agreement with the experiment is remarkable: bulk and surface related dispersions are captured not only qualitatively but also quantitatively. The difference between the spectra taken with the right- and left-circularly polarised light (Fig. \ref{fig3}h) allows us to identify additional features in the intensity distribution, making the agreement with the theory even better.

In spite of the clear correspondence between laser ARPES data and DFT calculations there is one detail which remains unexplained --- the striking flatness of the surface band without any signature of the Fermi level crossings.

\section{Superconductivity at the surface}\label{sec5}
\begin{figure}[h]%
\centering
\includegraphics[width=0.9\textwidth]{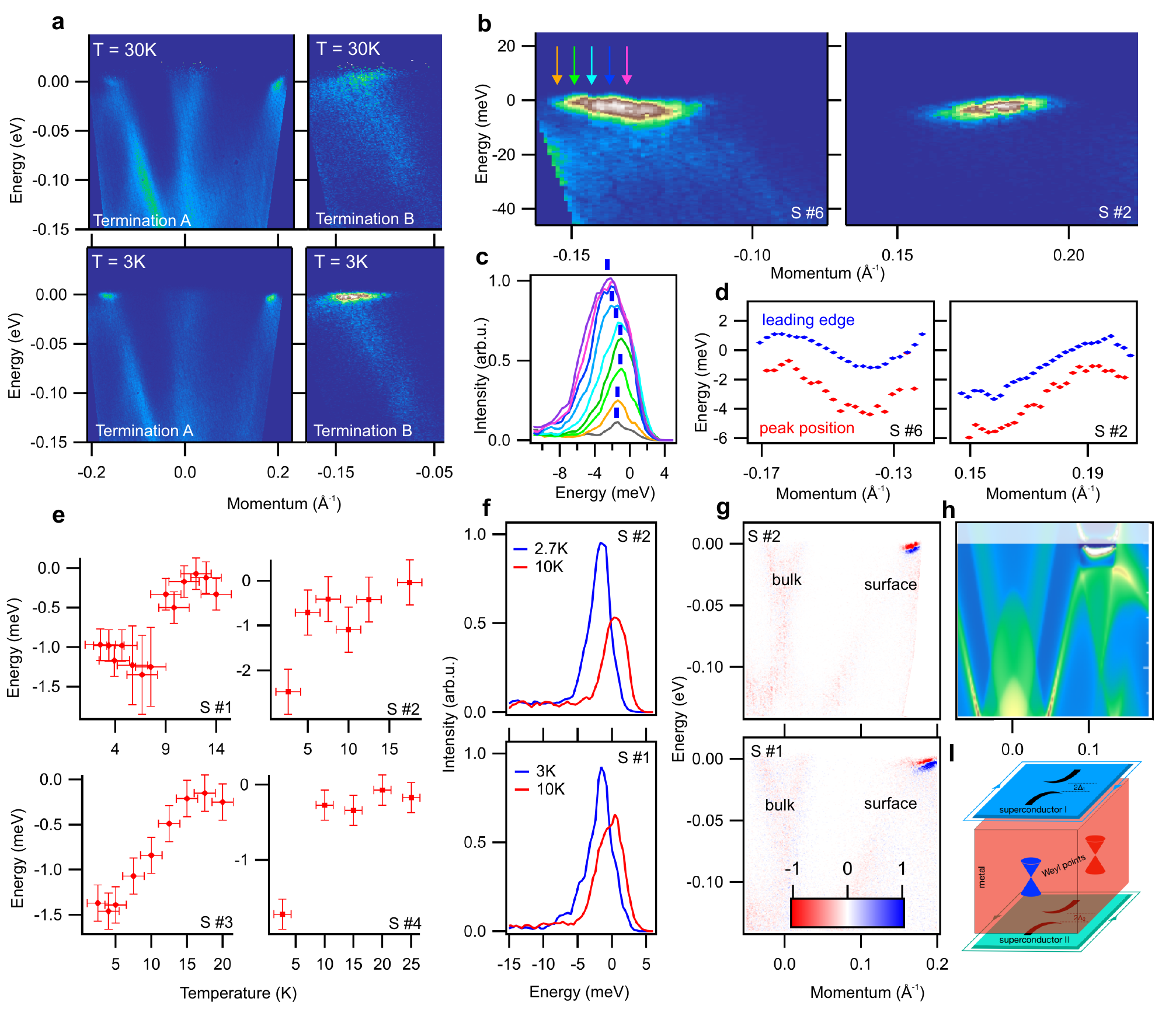}
\caption{Superconducting arcs. \textbf{a} Temperature dependence of the arcs dispersion from the terminations A and B. \textbf{b} Zoomed-in datasets showing underlying dispersion of the arcs. \textbf{c} EDCs corresponding to the coloured arrows in \textbf{b}. \textbf{d} Leading edge and peak positions from \textbf{b}. \textbf{e}  Averaged values of the peak positions closest to the Fermi level as a function of temperature for different samples and terminations. Left panels correspond to termination A, right panels - to termination B. \textbf{f} Shift of the EDCs with temperature. \textbf{g} Difference plots showing the changes of the intensity as a function of temperature. \textbf{h} Results of the calculated spectral weight taking into account the superconductivity at the surface. \textbf{i} Schematics of the electronic structure of PtBi$_2$. Green contours represent the Majorana states suggested by the topogical superconductivity at the surfaces.   }
\label{fig4}
\end{figure}
Record-high sharpness of the arc EDCs strongly resembles coherence peaks in ARPES data from superconductors (e.g. \cite{Kushnirenko_LiFeAs}). In order to find out whether the electronic states in question bear any other characteristic features of superconductivity, we have carried out temperature dependent measurements. In Fig. \ref{fig4}a we show the datasets recorded at 3 and 30K for both terminations. The comparison of the spectra taken at different temperatures underlines their flatness at lowest temperature. The arcs clearly lose spectral weight and gain dispersion - just as is to be expected when the system enters the normal state. The apparent asymmetry of the arcs dispersion stems from the openness of the Fermi contour made by an arc - conventional electron-like Fermi surface pockets would be supported by the symmetric with respect to the bottom dispersions crossing the Fermi level.

We have also reproducibly observed another peculiar for the superconducting state behavior. In Fig. \ref{fig4}b, where the arcs are measured with the highest resolution, the typical back-bending of the dispersion from the side where the states most closely approach the Fermi level is clearly seen. This is illustrated in panels c and d of Fig. \ref{fig4}, where we plot EDCs at several momentum values as well as their peaks and leading edge positions. We further track the behaviour of the peak positions as a function of temperature in Fig. \ref{fig4}e. Typical for superconductivity shifts are observed when going through the critical temperatures $T_c$. Such shifts measured at the $k_F$ give rather precise estimate of the superconducting gap $\Delta$. Our measurements yield $T_{cA} = 14\pm2$K and $T_{cB} = 8\pm2$ K whereas the corresponding superconducting energy gaps are equal to $1.4\pm0.2$ meV and $2\pm0.2$ meV respectively. The transition for the termination A seems to be broader, and $T_c$ higher compared to the surface B, which suggests that slightly differing superconducting states set in on the opposing surfaces. Taking into account the different electronic structure of the two surfaces with Fermi arcs it is not surprising that the superconducting orders are not fully equivalent. Signatures of Berezinskii-Kosterlitz-Thoules transition seen by transport \cite{Veyrat_Nanoletters} may explain the unusual $\Delta/T_c $ ratios.  

In panels f and g of Fig. \ref{fig4} we present additional evidence for essentially surface-related superconductivity observed in two different samples. While the EDC's corresponding to the surface states are clearly shifted upon varying the temperature (Fig. \ref{fig4}f), the bulk-related spectral weight remains virtually intact, showing only weak changes caused by slightly different width of the Fermi function. This is illustrated with the aid of 2D difference plots (Fig. \ref{fig4}g), where the clearly stronger variations of spectral function occur in the region where the arcs are located.

We can reproduce the experimental spectral function of PtBi$_2$ including both the flatness and back-bending of the surface states by switching on super-conductivity only at the surface via a solution of the Bogoliubov-de Gennes Hamiltonian for a semi-infinite solid with a gap function of $V_0 = 2$ meV in the first 3 PtBi$_2$ layers. 
The result is shown in Fig. \ref{fig4}h for the momentum and energy intervals corresponding to Fig. \ref{fig4}g (see also Fig.  S4 in SI). Note, that only the electron-electron part of the BdG spectral density is plotted to model the ARPES signal and that only the states around the Fermi arcs acquire a gap at the Fermi level.

The superconductivity of arcs in PtBi$_2$ follows not only from the emergence of unusually strong and sharp coherence peaks at low temperatures, flatness and back-bending of the dispersion as well as characteristic shifts of the EDCs. It follows also from the striking agreement with recent STM data \cite{Schimmel_gap} --- results of another surface-sensitive experiment on the crystals from the same batch. There (see Fig. 3 in \cite{Schimmel_gap}) the authors observed typical tunneling conductance of superconductors characterized by the superconducting gaps varying in space. Remarkably, the average value of the gap closely corresponds to the gap values determined by ARPES. Moreover, rather unusual considerable zero-biased conductance observed by Schimmel \textit{et al}\cite{Schimmel_gap} has now a very natural explanation in terms of a bulk contribution which remains ungapped. As can be seen from the ARPES data, the integrated contribution from the states associated with the bulk can easily reach a noticeable fraction of the signal from the surface, despite the dominant intensity of the arcs. The spot size of the laser beam in our study is of the order of 0.1 mm. This explains the agreement between the determined gap values with averaged STM data and does not exclude the possible existence of higher $T_c$ regions, the detection of which by ARPES would require an application of micro- or nano- variations of the technique.

PtBi$_2$ emerges as a stoichiometric Weyl semimetal with superconducting surfaces (Fig.\ref{fig4}i) and opens up a plethora of possibilities to manipulate topological and superconducting phases in a single material. For instance,  by varying the thickness of the single crystal, one can obtain a tunable intrinsic Josephson junction. Topological superconductivity at the surface should generate Majorana states at the edges. We tentatively assign the momentum-independent spectral weight at the Fermi level (apparently seen e.g. in Fig. \ref{fig2}d), enhanced when probing the regions of the surface with a number of terraces, to their presence. Further studies are needed to unambiguously identify and control both the high-$T_c$ superconductivity and Majorana fermions at the surfaces and edges of PtBi$_2$ single crystals and nano-structures.

\section{Methods}\label{sec7}
PtBi$_2$ single crystals with size up to 1cm were grown via self-flux method. For the ARPES study best crystals with dimensions 5 $\times$ 5 mm were selected. An important and convenient for ARPES experiments property of the single crystals is apparent absence of the stacking faults resulting in essentially the same and only termination upon multiple cleaving of the given sample.

ARPES measurements were carried out on the $1^2$- and $1^3$-ARPES end-stations at BESSY (HZB), as well as in the IFW laboratory using 5.9 eV laser light source. Samples were cleaved \textit{in situ} at a pressure lower than $1\times 10^{-10}$ mbar and measured at the temperatures of 15K and 1.5K at BESSY, and 3-30 K in the laboratory. The experimental data were obtained using the synchrotron light in the photon energy range from 15 to 50 eV with horizontal polarization and laser light with horizontal and circular polarizations . Angular resolution was set to 0.2$^{\circ}$ - 0.5$^{\circ}$ and energy resolution - to 2 - 20 meV. The findings from the experiments were consistent and reproducible across multiple samples.

The simultaneous presence of bulk non-superconducting and surface superconducting states makes the detection of true coherence peaks in ARPES challenging. Our experiments at the synchrotron with the energy resolution of the order of 5 meV turned out to be not sufficient to detect even the shifts of the leading edges of the corresponding arc peaks having FWHM of the order of 10meV and peak-to-background ratio of approximately 5. This is because the arc states are always located on top of the bulk continuum. Only measuring with energy resolution of the order of 1-2 meV did we manage to observe sufficiently sharp peaks (Fig. \ref{fig3} c,d and Fig. S3) and their sensitivity to the temperature. The sharpest features need to be found on the surface.  

Superconducting gap on the arcs is most likely anisotropic. We included to error bars in Fig. \ref{fig4}e the possible influence of a small shift of the beam spot and thus slightly different emission angle. Taking into account the very high localization in momentum space, this could potentially lead to probing a different part of the arc and thus different $k_F$, where the superconducting gap is slightly different.

We performed DFT calculations using FPLO\cite{KLAUS_FPLO}  within the general gradient approxiamtion (GGA)\cite{per96} and extracted a Wannier function model. This allows the determination of bulk projected spectral
densities (without surface states) and the spectral densities of semi-infinite slabs via Green's function techniques\cite{Sancho1985}. To model surface superconductivity of the semi-infinite slab the Wannier model is
extended into the Bogoliubov-de Genne formualism with a zero gap function in general except for a constant Wannier orbital diagonal singlet gap function matrix at the first 3 PtBi$_2$ layers. A modification of the Green's function method is used to accomodate this surface specific term. Further details are found in the supplementary material.

\section{Acknowledgements}\label{sec7}
We thank the Helmholtz-Zentrum Berlin f\"ur Materialien und Energie for for the allocation of synchrotron radiation beamtime. This work was supported withing the Collaborative Research Center SFB 1143 “Correlated Magnetism—From Frustration to Topology” and by the Dresden-W\"urzburg Cluster of Excellence (EXC 2147) “ct.qmat—Complexity and Topology in Quantum Matter”. SA acknowledges DFG via AS 523/4-1. S.A. and B.B. also acknowledge the support of DFG through Project No. 405940956. JIF acknowledges the support of the Alexander von Humboldt Foundation
via the Georg Forster Return Fellowship and ANPCyT grants PICT 2018/01509 and PICT 2019/00371. We thank Ulrike Nitzsche for technical assistance.

\section{Supplementary information}\label{sec7}
Here we present additional experimental and theoretical datasets which support our conclusions drawn in the main text.

In Fig. S1 we show ARPES Fermi surface maps obtained using the photon energies from 15 eV to 43 eV. Relatively strong variation of the pattern suggests a reasonable $k_z$-sensitivity of our experiment. We found the optimal value of inner potential to be equal to 10.5 V. This is in agreement with the previous study of Jiang et al. \cite{Jiang_JAP2020}

In Fig. S2 we show analog of Fig. 1 e-g, but here we compare experimental data with the results of band structure calculations carried out using the linear muffin-tin orbital (LMTO) method in the atomic sphere approximation (ASA) as implemented in PY
LMTO computer code. As is seen from the figure, the agreement is at the same level as earlier, underpinning the previous conclusion as regards the good agreement between experimental and theoretical 3D band structure. 

To model a system which has a non-zero gap-function only at the surface - in the first 30a$_B$ which is 3(PtBi$_2$) layers - we modified the standard Green\textquoteright s function technique for semi infinite slabs.
The system is build by a semi-infinite chain of identical blocks consisting of 3(PtBi$_2$) layers, repeating indefinitely away from the surface. Each block has a Hamiltonian $H_{k}$ for each pseudo momentum $k$
in the plane perpendicular to the surface and a hopping matrix $V_{k}$, which couples neighboring blocks. The blocks' minimum size is determined by the condition that $H$ and $V$ describe all possible hoppings.
In order to add superconductivity the Bogoliubov de Genne (BdG) formalism is used by extending the matrices in the following way
\begin{eqnarray*}
H_{k,\mathrm{BdG}} & = & \left(\begin{array}{cc}
H_{k} & \Delta_{k}\\
\Delta_{k}^{+} & -H_{-k}^{*}
\end{array}\right),\\
V_{k,\mathrm{BdG}} & = & \left(\begin{array}{cc}
V_{k} & 0\\
0 & -V_{-k}^{*}
\end{array}\right),
\end{eqnarray*}
where we choose $\Delta_{k}=\delta_{ii^{\prime}}\left(\begin{array}{cc}
0 & V_{0}\\
-V_{0} & 0
\end{array}\right)$ with $i$ being a spinless Wannier function (WF) index and the $2\times2$-matrix
to act in a single WFs spin subspace. This choice also leads to $\Delta\left[V_{k,\mathrm{BdG}}\right]=0$,
since $V$ is an off-diagonal part of the full Hamiltonian. To model
surface-only superconductivity we let $V_{0}=0$ for all (infinite)
blocks, except the first one, which gets a finite $V_{0}=2\mathrm{meV}$.

The standard Green's function solution for this problem consists of
determining the propagator $X$ which encompasses all diagrams, which
describe paths that start at a certain block, propagate anywhere towards
the infinite side of that block and returning to that block. $X$
also describes the Green's function $G_{00}$ of the first block as
well as the self energy to be added to the Hamiltonian to obtain $G_{00}$
(a self consistency condition) $G_{00}=X=\left(\omega^{+}-H-\Sigma\right)^{-1}$,
$\Sigma=VXV^{T}$ (in practice self consistency is obtained by an
accelerated algorithm, though). From this recursion relations can
be used to calculate all other Greens function blocks. These can be
derived by subdividing propagation diagrams into irreducible parts
using known components, in particular $X$.

If the first block differs from all the others (as is the case due
to $\Delta_{k}$) one needs to modify the method in the following
way. Let the first block have Hamiltonian $h$ and hoppings to the
second block $v$ (while all other blocks are described by $H$ and
$V$). Then the irreducible subdivision of the propagation diagrams
for $G_{00}$ results in ($g=\left(\omega^{+}-h\right)^{-1}$)
\begin{eqnarray*}
G_{00} & = & g+gvXv^{+}g+\left(gvXv^{+}\right)g\\
 & = & \frac{1}{\omega^{+}-h-vXv^{+}}
\end{eqnarray*}
which contains the surface Hamiltonian and a modified self energy
depending on the $X$ of the unmodifed semi-infinite slab. From this
we can derive the second's block Green's function
\[
G_{11}=X+Xv^{+}G_{00}vX
\]
and all others
\[
G_{n+1,n+1}=X+XV^{+}G_{nn}VX,\quad n>0
\]

which can be used to obtain the spectral density up to a certain penetration
depth. Note, that in our BdG case $H=H_{k,\mathrm{BdG}}\left[V_{0}=0\right]$,
$V=V_{k,\mathrm{BdG}}\left[V_{0}=0\right]$ and $h=H_{k,\mathrm{BdG}}\left[V_{0}\ne0\right]$,
$v=V$. The BdG spectral density is particle-hole symmetric and in
order to obtain results, which resemble ARPES data, one needs to use
the particle-particle block $G^{\mathrm{ee}}$ (the upper left quarter
of the $G$ matrix) only.

Fig. S4b shows the EDC of this method along the path denoted in Fig.
S4a. Note that a gap is opened at the surface band pockets close to
the Fermi energy, while the rest of the spectrum stays gap-less (if
we let $V_{0}\ne0$ for all blocks we get a completely gapped spectrum).
Fig. S4c shows a zoomed-in region around the surface state. Note,
that the bulk bands are gap-less (dark blue vertical features) while
the surface state shows a gap and corresponding band-back-bending.
The particle-hole symmetry becomes apparent, although with a larger
spectral weight for the occupied part due to the fact that we use
$G^{\mathrm{ee}}$ only.

\renewcommand{\thefigure}{S\arabic{figure}}
\setcounter{figure}{0}   
\begin{figure}[h]%
\centering
\includegraphics[width=0.9\textwidth]{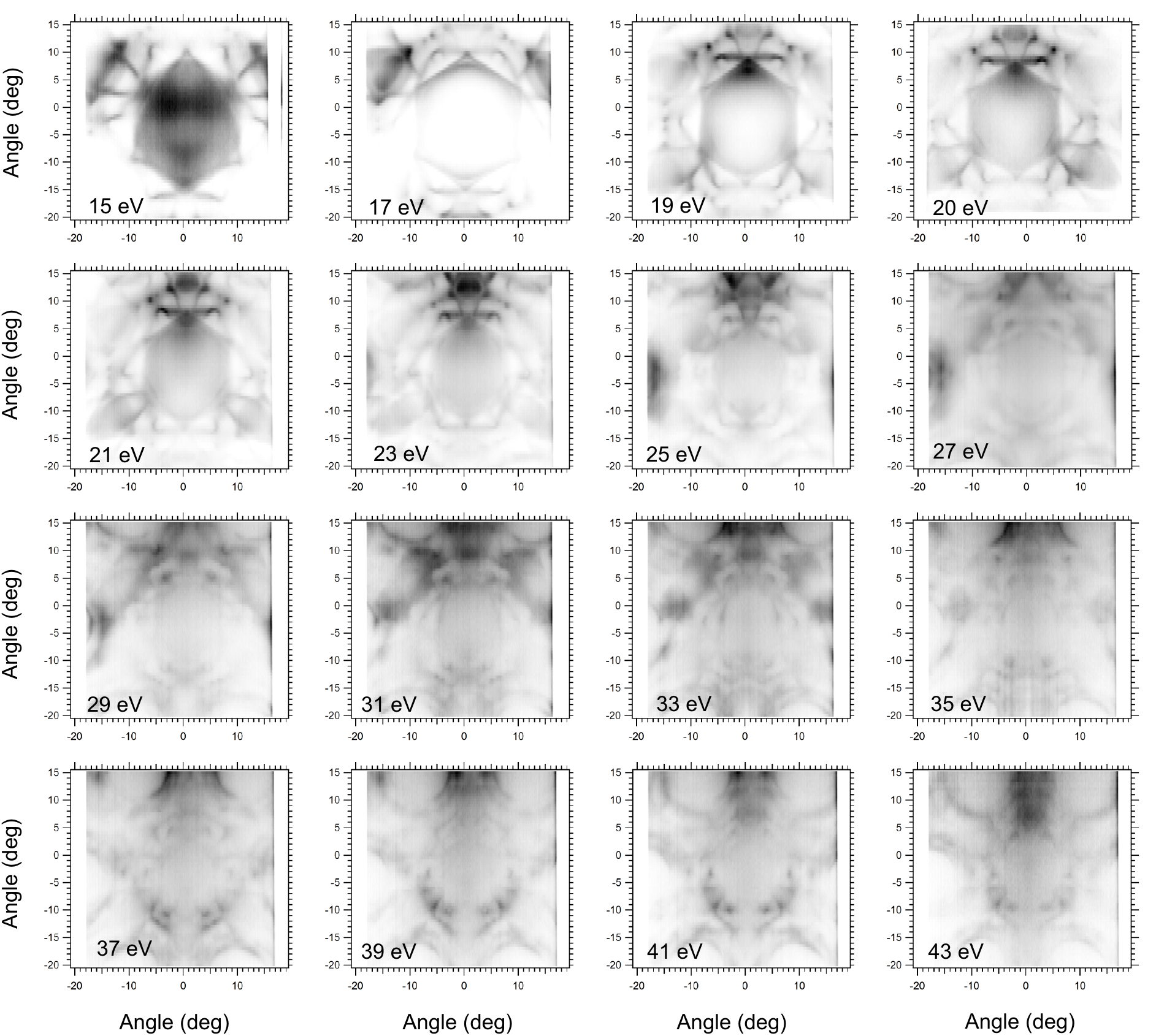}
\caption{Fermi surface maps from the 16 datasets taken using different photon energies.  }
\label{figS1}
\end{figure}

\begin{figure}[h]%
\centering
\includegraphics[width=0.9\textwidth]{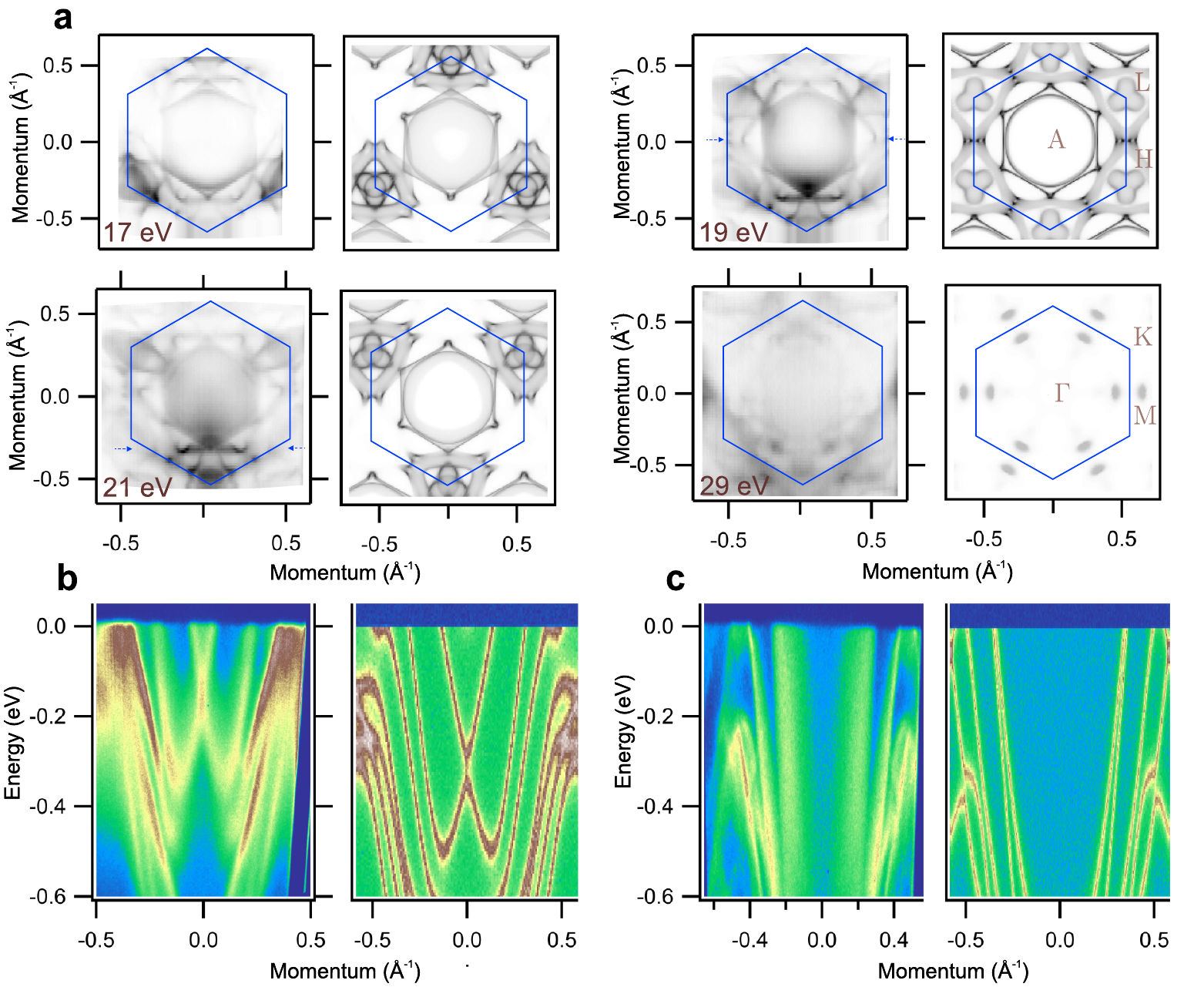}
\caption{\textbf{a}, Fermi surface maps taken using different photon energies and corresponding results of the band structure calculations. We note, that fixed photon energy probes a sphere of the large radius in the k-space, matching theoretical data formally only at one point in the center. \textbf{b}, \textbf{c} Energy-momentum intensity distributions at 21 eV and 19 eV respectively along the cuts indicated by blue dashed arrows in panel \textbf{e}.  }
\label{figS2}
\end{figure}

In Fig. S3 we present the sharpest EDCs from different samples and cleaves. Most of them have FWHM below 3 meV and peak-to-background ration of more than 30.

\begin{figure}[h]%
\centering
\includegraphics[width=1.0\textwidth]{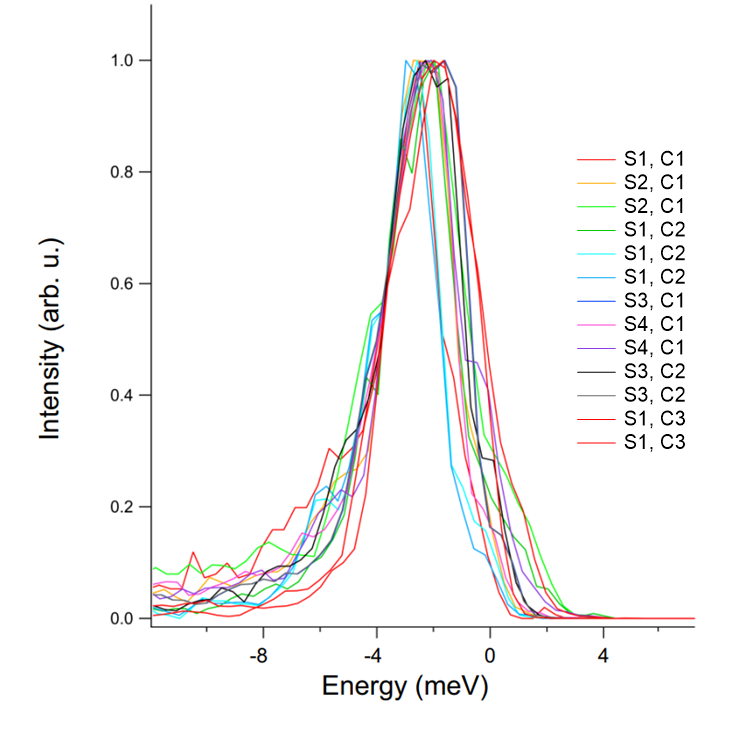}
\caption{Energy distribution curves taken close to $k_F$ of the arcs for different $k$, samples (S) and cleaves (C)  }
\label{figS3}
\end{figure}

\begin{figure}[h]%
\centering
\includegraphics[width=1.0\textwidth]{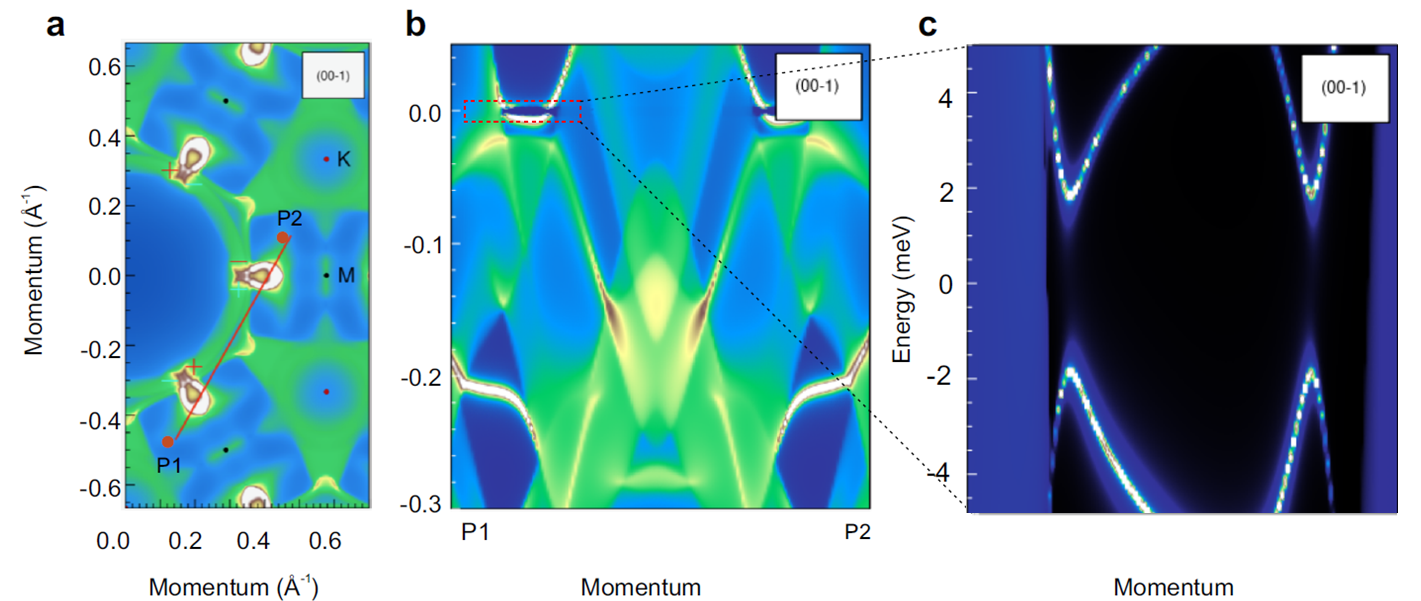}
\caption{\textbf{a} The path in the BZ. \textbf{b} The surface-only superconducting spectral function of the (00-1)-surface for $\Delta$ = 2meV and penetration depth 30a$_B$ ($G_{ee}$ only). Note that the gap is only open around the surface state pocket. \textbf{c} the blow-up of this pocket. Note, that the gap is open for the surface state but closed for the bulk bands   }
\label{figS4}
\end{figure}


\bibliography{sn-bibliography}% common bib file


\end{document}